\documentclass[aps,prb,twocolumn,floatfix,superscriptaddress]{revtex4-2}
\usepackage{graphicx,amsmath,amssymb,amsfonts,physics,dsfont}
\usepackage{multirow}
\usepackage{lipsum,color,colortbl}
\usepackage{verbatim,braket,booktabs}
\usepackage{empheq}

\usepackage{xcolor}

\definecolor{lightred}{RGB}{251,154,153}
\definecolor{lightgreen}{RGB}{178,223,138}
\definecolor{darkblue}{RGB}{31,120,180}
\definecolor{darkgreen}{RGB}{51,160,44}
\definecolor{lightblue}{RGB}{166,206,227}
\definecolor{darkred}{RGB}{227,26,28}
\definecolor{lila}{RGB}{106,61,154}
\definecolor{darkviolet}{RGB}{106,61,154}

\usepackage{graphicx}% Include figure files
\usepackage{dcolumn}% Align table columns on decimal point
\usepackage{bm}% bold math
\usepackage{hyperref}
\hypersetup{colorlinks=true, linkcolor=darkviolet, citecolor=darkgreen, urlcolor=darkblue}

\usepackage{tikz}

\usepackage{amsthm}
\usepackage{amsmath, amssymb}
\usepackage{physics}
\usepackage{booktabs}
\usepackage{url}

\renewcommand{\eprint}[2][]{\href{https://arxiv.org/abs/#2}{arXiv:~\nolinkurl{#2}}}

\begin{document}

\newcommand{\paul}[1]{\textcolor{red}{PF: #1}}
\newcommand{\luisa}[1]{\textcolor{darkgreen}{LE: #1}}
\newcommand{\zz}{$\mathbb{Z}_2$\,$\times$\,$\mathbb{Z}_2$ }

%Title of paper
\title{Critical lines and ordered phases in a Rydberg-blockade ladder}

\author{Luisa Eck}
%\email[]{luisa.eck@physics.ox.ac.uk}
\affiliation{Rudolf Peierls Centre for Theoretical Physics, University of Oxford, Parks Road, Oxford, OX1 3PU, United Kingdom}
\author{Paul Fendley}
\affiliation{Rudolf Peierls Centre for Theoretical Physics, University of Oxford, Parks Road, Oxford, OX1 3PU, United Kingdom}
\affiliation{All Souls College, University of Oxford, Oxford, OX1 4AL, United Kingdom}

\date{\today}

\begin{abstract}
Arrays of Rydberg atoms in the blockade regime realize a wealth of strongly correlated quantum physics, but theoretical analysis beyond the chain is rather difficult. Here we study a tractable model of Rydberg-blockade atoms on the square ladder with a $\mathbb{Z}_2$\,$\times$\,$\mathbb{Z}_2$ symmetry and at most one excited atom per square. We find $D_4$, $\mathbb{Z}_2$ and $\mathbb{Z}_3$ density-wave phases separated by critical and first-order quantum phase transitions. A non-invertible remnant of $U(1)$ symmetry applies to our full three-parameter space of couplings, and its presence results in a larger critical region as well as two distinct $\mathbb{Z}_3$-broken phases.  Along an integrable line of couplings, the model exhibits a self-duality that is spontaneously broken along a first-order transition.  Aided by numerical results, perturbation theory and conformal field theory, we also find critical Ising$^2$ and three-state Potts transitions, and provide good evidence that the latter can be chiral.
\end{abstract}

\maketitle

\section{Introduction}

Rydberg atoms provide an attractive platform to simulate strongly correlated quantum many-body systems because of their long lifetime and strong sensitivity to electric fields \cite{Bernien2017,Browaeys2020,Ebadi2021}. 
Neutral atoms in the ground state are trapped in optical tweezer arrays, and coupled to the highly excited Rydberg state with laser light. Strong dipole-dipole interactions can be tuned to forbid simultaneously exciting two atoms within a ``blockade'' radius.  The interactions and chemical potential can be tuned along with the blockade constraint, giving rise to a variety of interesting phases and transitions.

The Rydberg-blockade chain with nearest-neighbor occupancy forbidden displays some striking behavior \cite{Fendley2003}. With no other interactions, it exhibits quantum scars and the ensuing revivals \cite{Bernien2017,Turner2018}. Tuning a chemical potential gives an Ising transition to $\mathbb{Z}_2$ density-wave order.  Including strong attractive next-nearest-neighbor interactions eventually turns this transition into an integrable first-order line at a tricritical point. Making them strongly repulsive yields a $\mathbb{Z}_3$-ordered phase, with transitions out of it taking place via the three-state Potts conformal field theory (CFT), a chiral-clock transition, or an intermediate incommensurate phase \cite{Fendley2003,samajdar2018,chepiga2019,giudici2019,yu2022fidelity}. Longer-range interactions yield a $\mathbb{Z}_4$ ordered phase with Ashkin-Teller and chiral transitions \cite{Chepiga2021kibble,maceira2022conformal}.

The theoretical study of arrays beyond the chain is in its infancy, with most results thus far
stemming from numerics and bosonization. 
Rydberg atoms on square and triangle ladders exhibit cluster Luttinger liquid phases and supersymmetric critical points \cite{tsitsishvili2022phase,fromholz2022phase}, while Ising phase transitions occur by order-by-disorder mechanisms  \cite{sarkar2023}. A prominent recent proposal is to realize the 2d toric-code spin-liquid phase \cite{Verresen2020,Semeghini2021}, and coupling multiple Rydberg chains tuned to their Ising transition together can give rise to this physics \cite{slagle2022quantum}. 

In this paper we introduce and analyse a tractable model of Rydberg atoms on a square ladder. We impose a blockade allowing only one excitation per square, so the blockade radius is $\sqrt{2} <R_b <2$ in lattice units:
%\vspace{-0.05cm}
\begin{center}
  \begin{tikzpicture}[scale = 0.48]
   \clip  (-1.1, -0.45) rectangle (5.8, 1.35);
  \fill[fill=lightred!40!white, draw=black] (1,1) circle (1.55cm);
    \filldraw[lightred!40!white, draw=black] (3,0) circle (1.55cm);
    \draw[black] (1,1) circle (1.55cm);
    \foreach \x in {-1,...,5}
     { \foreach \y in {0,...,1}
        {
        \fill[gray] (\x,\y) circle (1mm);
        }
     }
     \fill[darkred] (1,1) circle (1.2mm);
     \fill[darkred] (3,0) circle (1.2mm);
     \draw[-latex] (1,1) -- (-0.4, 0.4);
     \node[below] at (0.6, 0.8) {\small $R_b$};
  \end{tikzpicture}
\end{center}
\vspace{-0.15cm}
We require a \zz symmetry and find a rich phase diagram that includes $D_4$, $\mathbb{Z}_3$ and $\mathbb{Z}_2$ density-wave orderings. Transitions are both first-order and critical, with the latter characterised by a free-boson orbifold, Ising$^2$ and Potts conformal field theories (CFTs), and a chiral-clock model. Using integrability, CFT and various simple limits along with DMRG and exact diagonalization, we map out the three-parameter phase diagrams displayed in Figs.~\ref{fig:heatmaps}, \ref{fig:diagramw0}, \ref{fig:phasediagramstDelta}. The non-invertible symmetry and self-duality described in our companion paper \cite{fendley2023pentagon} prove crucial. Their presence extends critical regions, as well as helping us to locate them precisely. The symmetry also allows us to characterize two distinct $\mathbb{Z}_3$-ordered phases, as it is spontaneously broken in only one.

In section \ref{sec:Hamiltonian}, we introduce our specific square-ladder Hamiltonian and its symmetries. Various extreme limits are analysed in section \ref{sec:limits}, yielding a variety of ordered phases as well as the transitions between them. In section \ref{sec:integrability}, we exploit the integrability along a particular line of couplings to understand the behaviour in the middle of the phase diagram.  In section \ref{sec:critical} we study two of the subtler transitions in more depth, allowing us to fully characterise the phase diagram. A review of the related Ashkin-Teller model and some perturbative results are collected in the appendices.

\section{Hamiltonian and symmetries}
\label{sec:Hamiltonian}

%\paragraph{The Hamiltonian}
We label a rung without excitations as the state $\ket{e}$, while excitations $\ket{t}$ and $\ket{b}$ on the top and bottom of rung $j$ annihilated by bosonic operators $t_j$ and $b_j$ respectively. The corresponding number operators are $n^t_j=t^\dagger_jt_j$ and $n^b_j=b^\dagger_j b_j$. Requiring at most one excitation per square means imposing  
\begin{align}
n^t_j n^t_{j+1}\,=\,n^t_j n^b_{j+1}\,=\,n^b_j n^b_{j+1}\,=0
\end{align}
for all $j$.
For $L$ rungs, the ensuing Hilbert space is dimension $2^L+(-1)^L$.
We utilize the linear combinations 
\begin{align}
p_j= \tfrac{1}{\sqrt{2}} \left( t_j + b_j \right)\ , \quad m_j = \tfrac{1}{\sqrt{2}} \left( t_j - b_j \right)
\end{align}
and define $\ket{+}=p_j^\dagger\ket{e}$ and $\ket{-}=m_j^\dagger\ket{e}$. We analogously define $n_j^-=m^\dagger_jm_j$ and $p_j^-=p^\dagger_jp_j$ so that the density of empty rungs is
\begin{align}
n^e_j=1-n^t_j-n^b_j=1-n^+_j -n^-_j\ .
\end{align}

A one-parameter integrable Hamiltonian with this constraint has appeared in various guises \cite{Finch2013,Braylovskaya2016,aasen2020topological,Lootens2021,fendley2023pentagon}. We deform it by including two more parameters, yielding
\begin{equation}\begin{aligned}\label{HRyd}
    H = &\sum_{j=1}^L \Big( (1-w) \big( p_j + p^\dagger_j \big)  + (1+2w) s_{j-1} s_{j+1} \\[-0.3cm]
    &\quad + (\Delta - 2t)  n^-_j + (\Delta+t) \left(n^e_{j-1} - n^e_{j+1}\right)^2\Big),
\end{aligned}
\end{equation}
where $s_j \equiv n^t_j-n^b_j$ and we take periodic boundary conditions.  In the $\pm$ basis, the first term creates or annihilates $+$ bosons, the second swaps  $+\leftrightarrow -$ on a pair of sites, the third is a chemical potential for $-$ bosons, and the last  is a next-nearest-neighbor interaction. The model is integrable for any $\Delta$ when $w$\,=\,$t$\,=\,0, and possesses a self-duality along this line  \cite{fendley2023pentagon}. We review the physics of the integrable line at the beginning of section \ref{sec:integrability}.

%\paragraph{The symmetries}
For even $L$, $H$ has a $\mathbb{Z}_2\times \mathbb{Z}_2$ symmetry exchanging bosons between top and bottom for all even and for all odd rungs. The even generator is
\begin{align}\label{eq:rydZ2}
{F}_{{\rm even}} = \prod_{l=1}^{L/2} \big(n^e_{2l} + t^\dagger_{2l}b^{}_{2l} + b^\dagger_{2l}t^{}_{2l}\big)=\prod_{l=1}^{L/2} (-1)^{n^-_{2l}} 
\end{align}
The other generator ${F}_{{\rm odd}}$ is given by replacing $2l$ with $2l-1$ in \eqref{eq:rydZ2}.
The combination of this $\mathbb{Z}_2\times \mathbb{Z}_2$ symmetry and our blockade constraint makes $H$ also invariant under a non-invertible symmetry \cite{fendley2023pentagon} generated by
\begin{align}\label{Qdef}
        &\mathcal{Q} = \tfrac{1}{2} \left( 1 + F \right) \sum_{j=1}^L \sum_{k=0}^{L-1} \prod_{l=0}^{k} (-1)^{1 - n^-_{j+l}},
\end{align} 
where $F = F_\text{even}F_\text{odd}$. Along the integrable line it can be understood as a relic of the $U(1)$ symmetry of the related XXZ chain, but here it applies throughout the full three-parameter space of \eqref{HRyd}. As apparent from \eqref{Qdef}, $\mathcal{Q}$ is non-vanishing only on the half of the Hilbert space invariant under $F$.   

\begin{figure}[t]
\centering
\includegraphics[width=0.48\textwidth]{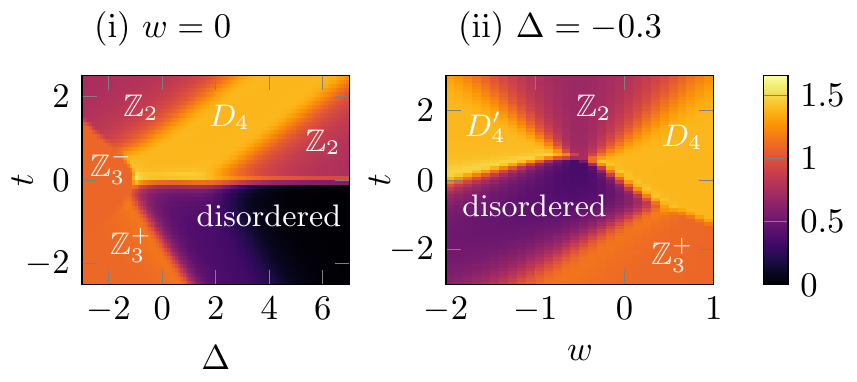}
\vspace{-0.2cm}
\caption{The half-cut ground-state entanglement entropy at $L$\,=\,12 from exact diagonalization for (i) $w$\,=\,0, (ii) $\Delta$\,=\,--0.3.}
\label{fig:heatmaps}
\end{figure}

\begin{figure}
    \centering
    \includegraphics{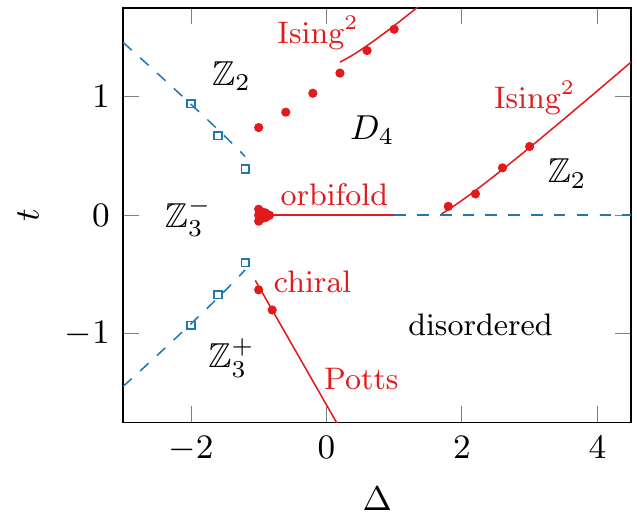}
    \vspace{-0.2cm}
    \caption{Phase diagram of the Rydberg ladder at $w$\,=\,0. \textcolor{darkred}{Critical (red filled circles)} and \textcolor{darkblue}{first-order (blue empty squares)} transitions are obtained with DMRG ($L$\,=\,121). The transition lines at $t$\,=\,0 are exact, while the others are perturbative. }
    \label{fig:diagramw0}
\end{figure}

A first look at the rich phase diagram is given in Fig.~\ref{fig:heatmaps},
showing the half-cut entanglement entropy for the ground state of $H$ along two planes within the three-parameter space of couplings. Here and elsewhere the exact diagonalization (ED) results are found using the package EDKit \cite{edkit}. 
Our more detailed results are summarized in the phase diagram given in Fig.~\ref{fig:diagramw0}. 

\section{Extreme limits}
\label{sec:limits}

We first analyse our Hamiltonian in various limits. All of the ordered phases are found in these limits, and several of the transitions are accurately determined using perturbation theory. We also explain in which region of couplings our ladder behaves effectively as the chain.

\subsection{The Ising$^2$ limit: $D_4$ and $\mathbb{Z}_2$ order}

We first take $\Delta+t$ positive and much larger than the other coefficients. The final term in \eqref{HRyd} dominates, and so all pairs of rungs $j$ and $j$\,+\,2 must either both be occupied by bosons, or both left empty. The even-$L$ Hilbert space then reduces to two decoupled sectors: one comprised of states $\ket{\pm e\pm e\pm \dots}$ and the other of states $\ket{e \pm e\pm e \dots}$. (The entirely empty state is higher energy and can be ignored.) In this limit, the effective Hamiltonian also decouples into two identical pieces:
\begin{align}\label{eq:Hisingsquared}
 \lim_{\Delta + t \to \infty}    H =\sum_{j=1}^L \left( (1+w) s_{j-1} s_{j+1} + (\Delta - 2t) n^-_j \right) . 
\end{align}
giving identical transverse-field Ising chains on odd sites in the first sector and even sites in the second. 

\begin{figure}
    \centering
    \includegraphics{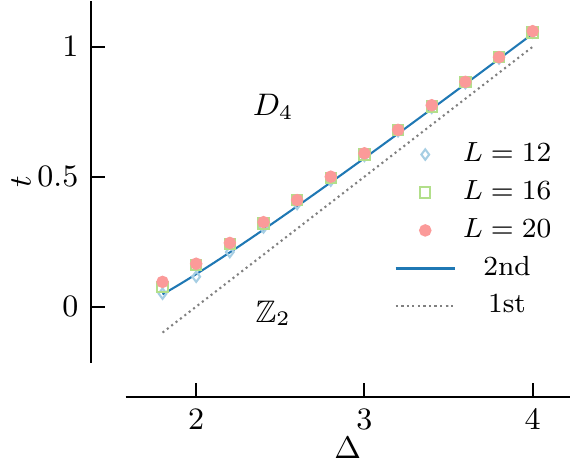}
        \caption{Numerical results (ED with \textcolor{lightblue}{$L=12$}, \textcolor{lightgreen}{$L=16$}, \textcolor{lightred}{$L=20$}) and analytical results (\textcolor{gray}{1st} and \textcolor{darkblue}{2nd} order perturbation theory \eqref{eq:pertresult} ) for the locations $t_+(w, \Delta)$ of Ising squared transitions between the $\mathbb{Z}_2$ ordered phase and the $D_4$ ordered phase.  }
    \label{fig:isingnew}
\end{figure}

Ancient results locate a critical point between order and disorder when the couplings of the two terms in \eqref{eq:Hisingsquared} are equal in magnitude, here corresponding to  $\Delta$\,--\,2$t$\,=\,$\pm 2(1+w)$ \cite{Onsager1943}. The $p_j$\,+\,$p_j^\dagger$ term in $H$ gives rise to corrections, which can be dealt with in perturbation theory, as described in Appendix \ref{app:pert}. Their effect is to renormalize the $n_j^-$ term in \eqref{eq:Hisingsquared}, shifting the transitions to
\begin{align}
t_{\pm}= \tfrac{\Delta}{2} \pm |1+w| + \tfrac{(1+w)^2}{6\Delta\pm 4|1+w|}+\mathcal{O}\big(\Delta\pm\tfrac{2|1+w|}{3}\big)^{-2}.
\label{eq:pertresult}
\end{align}
These curves are plotted in Figs.~\ref{fig:isingnew} and \ref{fig:diagramw0} along with data for the transition found from ED.  The analytical and numerical results agree well for large enough $\Delta$. The two sectors remain independent until the transitions break down, as it takes order $L$ actions of $H$ to mix them.  The continuum limit all along these transition lines is thus described by the Ising$^2$ CFT.

The symmetry-breaking patterns in this Ising$^2$ limit \eqref{eq:Hisingsquared} are easy to obtain.  The $\mathbb{Z}_2$ translation symmetry exchanging even and odd sites is spontaneously broken throughout this region.  The Ising-disordered regions $t>t_+$ and $t<t_-$ are therefore denoted by $\mathbb{Z}_2$ in Fig.~\ref{fig:diagramw0}.  For $t_-<t<t_+$ the Ising chains are ordered, and \zz exchange symmetry is broken as well. Since translation symmetry relates $F_{\rm even}$ and $F_{\rm odd}$, the full broken symmetry group in this phase is the dihedral group $D_4$, the symmetry group of the square. The four ground states depend on the sign of 1\,+\,$w$. For the Ising antiferromagnet $w$\,$>$\,--1,  as $t\to \infty$ in this region they are $\ket{tebeteb\dots}$ and its translations. For the Ising ferromagnet $w$\,$<$\,--1, as $t\to \infty$ they are $\ket{tetete\dots}$, $\ket{bebebe\dots}$ and translations. While the broken symmetry is $D_4$ in both cases, the two phases are distinguishable, as  translation symmetry relates all four ground states in the former phase but not the latter. We thus denote the latter phase with a prime, and it is visible at the left of Fig.~\ref{fig:heatmaps}(ii) as well as in Fig.~\ref{fig:phasediagramstDelta}(ii)-(iii) below.

\subsection{$\mathbb{Z}_3$ phases and their transitions}

The physics changes dramatically, not surprisingly, if we consider the opposite sign. Sending $\Delta+t$\,$\to$\,$-\infty$, we find two distinct $\mathbb{Z}_3$ phases.  When $\Delta - 2t$ remains finite, the final term of $H$ dominates and is minimized for a boson on every third site.  For $\Delta>2t\to-\infty$ and $L$ a multiple of 3, the three ground states approach $|$\,+\,$e\,e$\,+\,$e\,e\dots\rangle$ and its translations. For $2t>\Delta\to -\infty$, the three ground states are $|$\,--\,$e\,e$\,--\,$e\,e\dots\rangle$ and its translations. The latter are eigenstates of $H$ and so remain exact ground states throughout a region with $t$ and $\Delta$ finite.   Both regions thus have $\mathbb{Z}_3$ density-wave order, but for $L$ a multiple of 12, no other conventional symmetry is spontaneously broken. 

Remarkably, the two phases can still be distinguished. The non-invertible symmetry \eqref{Qdef} is spontaneously broken in the latter but not the former. Indeed, while $\mathcal{Q}|$\,+\,$e\,e$\,+\,$e\,e\dots\rangle$\,=\,0,  the exact ground states $|\dots$\,--\,$e\,e$\,--\,$e\,e\dots\rangle$ maximize this charge. We thus dub the two phases $\mathbb{Z}_3^+$ and $\mathbb{Z}_3^-$ respectively, and they are readily located in Fig.~\ref{fig:diagramw0} and Fig.~\ref{fig:phasediagramstDelta}(ii)-(iii). 

The first-order transition between the two $\mathbb{Z}_3$ phases occurs when their ground-state energies are equal. In this limit, the effective Hamiltonian is simply 
\begin{align} \label{eq:Z3Ham}
 \lim_{\Delta + t \to -\infty}    H =\sum_{j=1}^L (\Delta - 2t) n^-_j , 
\end{align}
so the transition occurs when its coefficient vanishes. Including perturbative corrections as described in appendix \ref{app:pert} locates the transition at  
\begin{align}
t_{3^-3^+}\approx \tfrac{\Delta}{2} -\tfrac{(1-w)^2}{6\Delta} + \mathcal{O}\big(\tfrac{1}{\Delta^2}\big) \ .
\label{t3minusplus}
\end{align}
As apparent from the bottom left of Fig.~\ref{fig:diagramw0}, the agreement between this curve and our DMRG numerics is good for a large region. We have no analytic results for the transitions between the $D_4$ and the $\mathbb{Z}_3$ phases, but our DMRG numerics indicate that they are first-order as well. The transition from the $\mathbb{Z}_3^+$ to the disordered phase is subtler, and we defer a detailed discussion to Section \ref{sec:chiral}.

The other transition out of the $\mathbb{Z}_3^-$ phase can be understood by still requiring $\Delta+t$\,$\to$\,$-\infty$ but now allowing for $\Delta-2t$\,$\to$\,$-\infty$ as well. The effective Hamiltonian is then comprised of the last two terms of $H$, which are diagonal in the $\pm$ basis. The $\mathbb{Z}_2$ phase also occurs in this limit, with ground states $|e$\,--\,$e$\,--\,$e$\,--$\dots\rangle$ and its translation. The first-order transition occurs when the ground state-energies of the $\mathbb{Z}_3^-$ and $\mathbb{Z}_2$ phases are equal. Including the perturbative corrections from the off-diagonal terms in $H$ (see Appendix \ref{app:pert}) locates it at
\begin{align}
t_{3^-2^-}\approx -\tfrac{\Delta}{2} +\tfrac{(1+2w)^2}{8\Delta} + \mathcal{O}\big(\tfrac{1}{\Delta^2}\big)\ .
\label{t3minus}
\end{align}
The numerically determined transitions follow the perturbative results (\ref{t3minusplus},\ref{t3minus}) fairly well, as apparent in Fig.~\ref{fig:diagramw0}.

%, as shown in Fig.~\ref{fig:diagramsnew}.

\subsection{The chain limit}
\label{sec:chain}

In the limit $\Delta-2t\to\infty$, the $-$ bosons are forbidden. The effective Hamiltonian describing the $+$ bosons is
\[H_+=\sum_{j=1}^L \Big( (1-w)(p_j + p_j^\dagger) %\\[-0.2cm]
  +2(\Delta + t) \big(n^+_j-n^+_{j-1}n^+_{j+1}\big) \Big). %, \quad n^e_{j-1} = n^e_{j+1} \forall j
\]
Ignoring the $-$ bosons makes $H_+$ precisely the Hamiltonian of the Rydberg-blockade chain along a line of couplings where the chain chemical potential $U$ is equal to $-V$, the next-nearest-neighbor interaction strength \cite{Fendley2003}. The precise correspondence is $U=-V=2(\Delta+t)/(1-w)$, where we set to 1 the annihilation/creation coefficient of \cite{Fendley2003}. This line includes the ``PXP'' model possessing quantum scars \cite{Turner2018} at $U$\,=\,$V$\,=\,0, which is $\Delta$\,=\,$-t$\ here. 

For large negative $\Delta+t$, $H_+$ is in the $\mathbb{Z}_3^+$ phase, with a critical transition to the disordered phase \cite{Fendley2003}. 
The chain results allow us to locate the analogous transition in the three-parameter space of the ladder. The $U$\,=\,$-V$ line of the chain comes close to the exact three-state Potts critical point at $U$\,$\approx$\,$-3.03$, $V$\,$\approx$\,3.33. Since the transition line is smooth as $U$ and $V$ are varied away from the Potts values, we can approximate the critical value on the nearby $U$\,=\,$-V$ line as $U$\,=\,$-V\approx$\,-3.2. The $\mathbb{Z}_3^+$ to disorder transition therefore occurs at 
\begin{align}\label{t3dis}
t_{3^+,\text{dis}} \approx -\Delta -1.6(1-w) + \mathcal{O}\big(\tfrac{1}{\Delta}\big)\ ,
\end{align} 
as apparent in Fig.~\ref{fig:diagramw0}. 

The $\mathbb{Z}_3$ to disordered transition in the chain can be in three-state Potts or chiral-clock universality classes, or take place via an intermediate incommensurate phase \cite{Fendley2003,samajdar2018,chepiga2019,giudici2019}. There the analysis is aided both by integrability and an extreme limit where incommensurability can be established analytically. Unfortunately, these avenues are not available in the square ladder, as the integrable line here does not cross this transition. We thus need to use numerics to understand the nature of this transition, a task we defer to section \ref{sec:chiral}.

\section{The integrable line}
\label{sec:integrability}

We obtain exact results in the region with no large couplings by exploiting the integrability of $H$ along the line $w$\,=\,$t$\,=0. As detailed in our companion paper \cite{fendley2023pentagon}, the spectrum here
can be mapped to that of the XXZ chain with a variety of boundary conditions, yielding three separate phases as $\Delta$ is tuned. The phase diagram along this line is
\vspace{-0.1cm}\begin{center}
    \begin{tikzpicture}[scale = 0.9]
    \begin{scope}[shift = {(0.,0)}]
        \fill[lightred] (0.8, -0.1) rectangle (4.2, 0.1);
        \draw[-latex, thick] (-1.8,0) -- (6.7,0) node[right] {$\Delta$};
        \draw[thick] (0.8, 0.15) -- (0.8, -0.15) node[below] {$\Delta$\,=\,$-1$};
        \draw[thick] (4.2, 0.15) -- (4.2, -0.15) node[below] {$\Delta$\,=\,1};
        \node[above, text=darkred, align=center] at (2.5,0.2) {\small free-boson\\ \small orbifold CFT}; 
  \node[above right, align=left, text=darkgreen] at (-1.4,0.1){broken $\mathcal{Q}$};
  \node[above left, text=darkblue, align=center] at (6.7,0.1) {\small broken\\\small self-duality};
      \end{scope}
    \end{tikzpicture}
\end{center}
\vspace{-0.1cm}
with a gapless phase for $|\Delta|\le 1$.
The $\mathbb{Z}_3^-$ phase holds for $\Delta<-1$, and we explain next how the rest of the line describes transitions to the disordered phase. 

\subsection{The $\mathbb{Z}_2$-disorder transition}

Three ground states $\ket{eeee\dots}$, $\ket{+\,e+e\dots}$ and $\ket{e+e+e\dots}$
coexist as $\Delta\to\infty$ along the integrable line. No conventional symmetry mixes all three, but for $w$\,=\,$t$\,=\,0 a non-invertible self-duality map does \cite{fendley2023pentagon}. Although the ground states change for $\Delta$ finite, the three remain degenerate for all $\Delta$\,$>$\,1 and spontaneously break the self-duality. This degeneracy is characteristic of a first-order transition between the disordered phase and the $\mathbb{Z}_2$ density-wave order apparent in the Ising$^2$ limit. 

\begin{figure}
    \centering
    \includegraphics{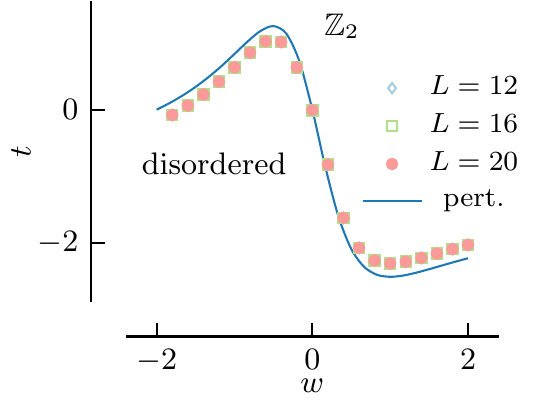}
    \vspace{-0.2cm}
    \caption{Numerical results from ED compared to the perturbative curve \eqref{eq:firstorderresult} for the location $t_{2,{\rm dis}}$ of the first-order transition between the $\mathbb{Z}_2$ and disordered phases at $\Delta = 2.5$}
    \label{fig:firstorder}
\end{figure}

This first-order transition, readily apparent in Fig.~\ref{fig:diagramw0}, cannot be located exactly away from the integrable line, but perturbation theory works well. For large $\Delta$, the diagonal terms dominate, and we find (see App.~\ref{app:pert})
\begin{equation}\label{eq:firstorderresult}
    t_{2,{\rm dis}} \approx -\Delta w\frac{2+w}{1+2w^2}\ .
 \end{equation}
We locate the first-order transition numerically by using ED to find when the gap between the lowest two momentum-zero states closes. The results together with the curve \eqref{eq:firstorderresult} are plotted in Fig.~\ref{fig:firstorder} for $\Delta=2.5$, with excellent agreement.

\subsection{Operators in the CFT}
\label{sec:CFT}

A non-invertible map to the XXZ chain \cite{Braylovskaya2016,Lootens2021,fendley2023pentagon} guarantees that $H$ is critical for $|\Delta|\le 1$ and $w$\,=\,$t$\,=0. Its continuum limit is described by a free-boson orbifold conformal field theory (other than at $\Delta\ne -1$, where the fermi velocity is zero). This $D_4$-invariant CFT is labeled by a ``radius'' $R$ and contains operators $C_{l,m}$ of dimension $x_{l,m}$ for non-negative integers $l,m$, where
\begin{align}
\label{CFTdim}
\Delta = -\cos\tfrac{2\pi}{9}R^2\ ,\qquad x_{l,m} = \tfrac{l^2}{4R^2} + m^2R^2
\end{align}
so that 0\,$<$\,$R$\,$\le\,$3/$\sqrt{2}$.
The other primary operators in the orbifold CFT are called ``twist'' fields. There are two of dimension $x=\frac18$ and two of dimension $\tfrac98$, denoted by $\sigma_{1,2}$ and $\sigma_{1,2}'$ respectively. 
The behavior of these operators under the $D_4$ and non-invertible symmetries is discussed in detail in \cite{fendley2023pentagon}.

Here we give more detail into how lattice operators in the integrable Rydberg ladder correspond to those in the CFT. Not only does this give useful intuition, but proves to be valuable in understanding how to perturb away from the integrable line. 
To do so, we exploit the symmetries of the lattice model and the CFT. However, translation symmetry is subtler. As explained in \cite{fendley2023pentagon}, the lattice analog of the $D_4$ symmetry of the CFT comes from combining the $\mathbb{Z}_2 \times \mathbb{Z}_2 $ symmetry with translation symmetry. This result is in harmony with our results in the Ising$^2$ limit here, where the broken $D_4$ indeed involves translation symmetry.

\begin{figure}
\begin{center}
\includegraphics[scale=1]{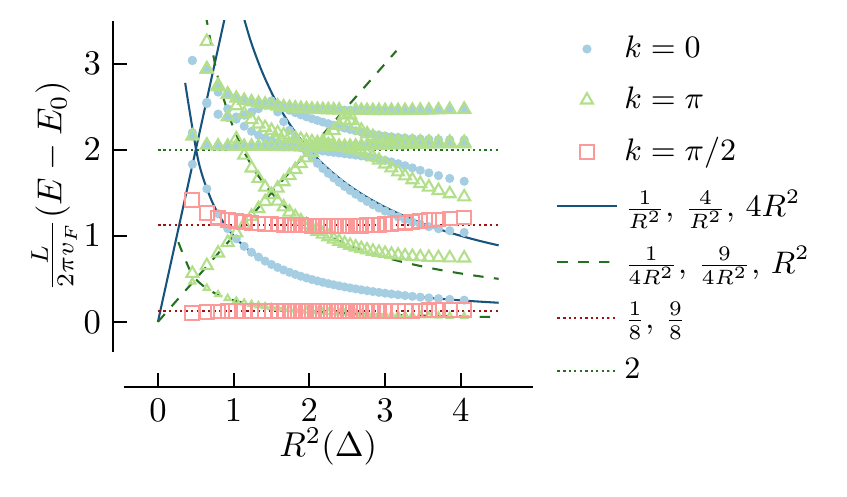}
\end{center}
 \vspace{-0.85cm}
\caption{\small
The lowest-lying rescaled energy gaps of the $L = 12$ integrable Rydberg ladder with parameter $\Delta$ found using ED are plotted against the squared radius $R^2(\Delta)$. }
\label{fig:Rsquares}
\end{figure}

To confirm \eqref{CFTdim} and to help us identify the symmetries, we first study the finite-size spectrum of $H$. Since the energies are related to the scaling dimensions \cite{Blote1986,Affleck1986}, a numerical determination of the former gives the latter. The precise relation valid as $L\to\infty$ is
\begin{align}
\label{Evf}
E= \frac{2\pi v_F}{L} \big(x - \tfrac1{12}\big)\ , \qquad  v_F=\pi \frac{\sqrt{1 - \Delta^2}}{\arccos{\Delta}}
\end{align}
where the fermi velocity $v_F$ is determined using the integrability of the XXZ chain.
We use ED to find the energies for a variety of $\Delta$, and give the results in Fig.~\ref{fig:Rsquares}. The agreement with the CFT dimensions is excellent. We also find that states created by ${C}_{l,m}$ have lattice translation eigenvalue $e^{ik}=(-1)^{l+m}$, while those created by the twist fields have $e^{ik} =\pm i$. The unit cell for taking the continuum limit is therefore comprised of four sites, in harmony with our results for the $D_4$ phase. We also have checked that a variety of levels of dimension $x_{l,m}$ in Fig.~\ref{fig:Rsquares} have charge $4m^2$ under the non-invertible operator ${\mathcal{Q}}$, as predicted in \cite{fendley2023pentagon}.

The simplest operators commuting with the $\mathbb{Z}_2 \times \mathbb{Z}_2 $ symmetry as well as $\mathcal{Q}$ are those in the Hamiltonian. The Hamiltonian along the integrable line is comprised of two self-dual operators, namely 
 \begin{align}\label{O1Delta}
    \widehat{O}^1_{j} = p^\dagger_j + p^\dagger_j + s_{j-1} s_{j+1}\ ,\quad 
    \widehat{O}^\Delta_{j} = n^-_j  + \left( n^e_{j-1} - n^e_{j+1} \right)^2\ .
\end{align}
The operators coupling to $w$ and $t$ in \eqref{HRyd} are chosen to be odd under the self-duality of the integrable line, and are
\begin{align}\label{Owt}
    \widehat{O}^w_j = 2s_{j-1} s_{j+1}-p_j - p^\dagger_j\ ,
    \widehat{O}^t_j =  \left( n^e_{j-1} - n^e_{j+1} \right)^2- 2n^-_j.
\end{align}
Since translation symmetry on the lattice becomes an internal symmetry of the CFT, it is useful to consider both staggered and unstaggered lattice operators. For each of these four operators we thus define the combinations
\begin{align} 
\widehat{O}^+_{j} \equiv \widehat{O}_j + \widehat{O}_{j+1}\ ,
\qquad \widehat{O}^-_{j} \equiv \widehat{O}_j - \widehat{O}_{j+1}\ . 
\end{align}

The swap operators 
 \begin{equation}\label{eq:flip}
    s_j =  p^\dagger_{j} m_{j} + m^\dagger_{j} p_{j},\qquad s'_j = p^\dagger_{j} m_{j} - m^\dagger_{j} p_{j}.
\end{equation}
provide another basic set of operators. These
are odd under the $\mathbb{Z}_2$ symmetries $F_\text{even}$ and $F_\text{odd}$ for $j$ even and odd respectively. Thus we do not take the staggered and unstaggered combinations here, but rather treat operators for $j$ even and odd separately. Indeed, the two-point function $\langle s_j s_k\rangle$ vanishes unless $j$ and $k$ are both even or both odd.

It is natural to identify the continuum limit of the simplest lattice operators with the simplest CFT fields possessing the same symmetries. Since the integrable Hamiltonian is comprised of the two self-dual operators from \eqref{O1Delta}, varying the coefficient of either changes $R$ in the CFT, or equivalently, the stiffness of the free boson field $\Phi$. The lattice operators $\widehat{O}^{1+}$ and $\widehat{O}^{\Delta+}$ should thus correspond to $|\nabla\Phi|^2$ in the CFT, the exactly marginal field.
The anti-self-dual operators defined in \eqref{Owt} are
invariant under both ${F}_\text{even}$, $F_\text{odd}$ and $\mathcal{Q}$. They therefore should correspond to ${C}_{1,0}$ and ${C}_{2,0}$ in the staggered and unstaggered cases respectively. The swap operators $s_{\rm even}$ and $s_{\rm odd}$ are odd under the corresponding $F_a$, and so should correspond to $\sigma_1$ and $\sigma_2$. Likewise, $s'_{\rm even}$ and $s'_{\rm odd}$ should correspond to $\sigma_1'$ and $\sigma_2'$. We summarize these correspondences in Table  \ref{tab:CFToperators}.
 \begin{table}[h]
    \centering
    \begin{tabular}{l l l l l l}
    \toprule
        lattice: & $(\widehat{O}^{1,\Delta})^+$\ &  $(\widehat{O}^{w,t})^-$\  & $(\widehat{O}^{w,t})^+$\ & $s_{j\,\text{even,odd}}$\ & $s'_{j\,\text{even,odd}}$ \\
    \midrule 
        CFT: & $|\nabla\Phi|^2$  & ${C}_{1,0}$  & ${C}_{2,0}$ & $\sigma_1,\,\sigma_2$ & $\tau_1,\tau_2$  \\[0.05cm] 
       dim:\ & $2$ & $\frac{1}{4R^2}$ & $\frac{1}{R^2}$ & $\frac{1}{8}$ & $\frac{9}{8}$ \\ 
    \bottomrule 
    \end{tabular}
    \caption{Correspondence between lattice and CFT operators}
    \label{tab:CFToperators}
\end{table}

\begin{figure*}
\begin{center}
\includegraphics[scale=1]{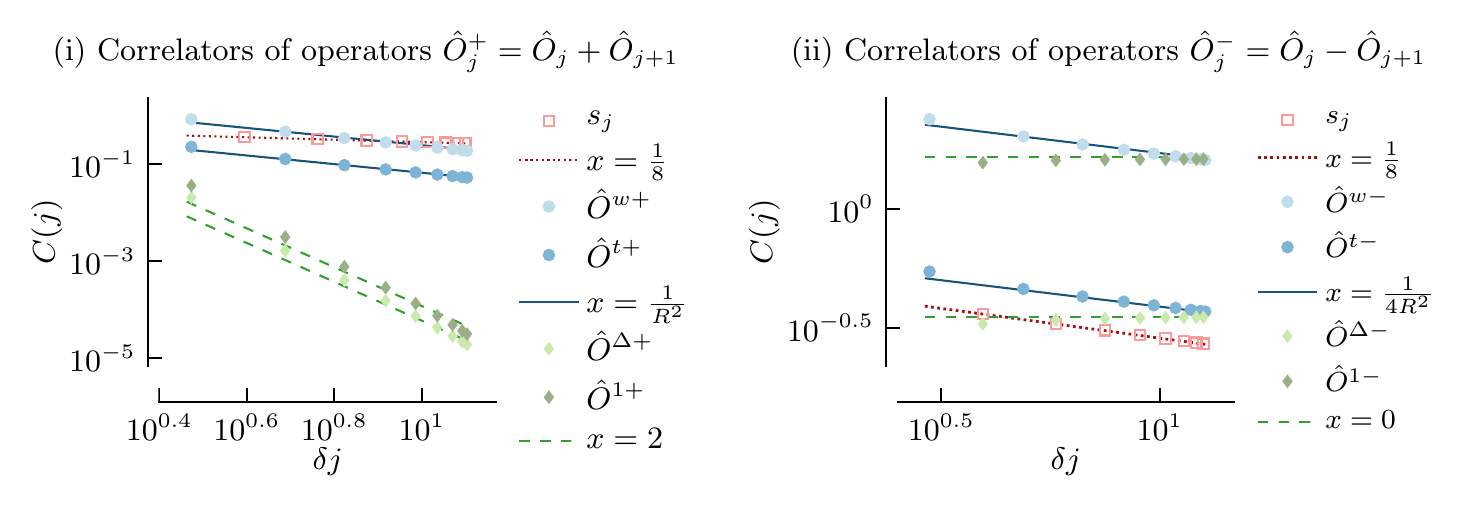}
\end{center}
 \vspace{-0.85cm}
\caption{\small
 The two-point correlator \eqref{eq:correlationfunctions} of the periodic $L$\,=\,40 integrable Rydberg ladder at $\Delta$\,=\,0 found using DMRG for several (i) unstaggered and (ii) staggered lattice operators. The slope is compared to the expected CFT scaling dimensions.}
\label{fig:correlationfunctions}\
\end{figure*}

To confirm these identifications, we  compute their two-point functions in the ground state.
The long-distance behavior of this correlation function in a critical theory is 
\begin{equation}\label{eq:correlationfunctions}
    \langle \widehat{O}_0 \widehat{O}_{j} \rangle  - \langle \widehat{O}_0 \rangle \langle \widehat{O}_j \rangle \sim c \big(\delta j)^{-2x_\mathcal{O}}\ ,\quad
\delta j \equiv   \tfrac{L}{\pi}  \sin{\left(\tfrac{\pi}{L} j\right)}\ ,
\end{equation}
where $x_\mathcal{O}$ is the scaling dimension of the corresponding continuum field $\mathcal{O}$ governing this behavior.
The correlation functions are obtained with DMRG using the ITensor library \cite{itensor}, and the hard-core nearest-neighbour constraint is implemented as described in \cite{giudici2019}. 
We plot some for $\Delta=0$ in Fig.~\ref{fig:correlationfunctions}. 
We collect our numerical results for a variety of lattice operators for various $\Delta$ in Table \ref{tab:rydoperators}, giving strong support to the correspondences in Table \ref{tab:CFToperators}.

 \setlength{\tabcolsep}{7pt}

\begin{table*}[t]
    \centering
      \small
    \begin{tabular}{l l l l l l}
    \toprule
%   & \multicolumn{5}{c}{measured scaling dimension at $L=40$, pbc} \\ 
 lattice operator\     & $\Delta = -0.5$ & $\Delta = -0.2$ & $\Delta = 0$ & $\Delta = 0.2$ & $\Delta = 0.5$ \\ %& $\Delta = 1.$ \\
    \midrule
    $\widehat{O}^{\Delta+}_ j$ & $2.30 \pm 0.20$ & $2.42 \pm 0.12$ & $2.44 \pm 0.16$ & $2.41 \pm 0.20$ & $2.34 \pm 0.19$ \\ %& $2.09 \pm 0.10$ \\ 
    $\widehat{O}^{1+}_{j}$ &  $2.37 \pm 0.22$ & $2.54 \pm 0.16$ & $2.56 \pm 0.26$ & $2.50 \pm 0.29$ & $2.39 \pm 0.28$ \\ % & $2.21 \pm 0.17$ \\ 
    \rowcolor{lightblue!50!white}  $|\nabla\phi|^2$ & 2 & 2 & 2 & 2 & 2 \\[0.05cm]
  $\widehat{O}^{w+}_{j}$ & $0.680 \pm 0.019$ & $0.523 \pm 0.015$ & $0.456 \pm 0.012$ & $0.404 \pm 0.011$ & $0.341 \pm 0.009$ \\ 
    $\widehat{O}^{t+}_{j}$ & $0.669 \pm 0.010$ & $0.514 \pm 0.007$ & $0.449 \pm 0.007$ & $0.399 \pm 0.008$ & $0.341 \pm 0.011$ \\
     $(n^+_j + n^-_j)^+$ & $0.717 \pm 0.001$ & $0.534 \pm 0.007$ & $0.461 \pm 0.007$ & $0.408 \pm 0.006$ &  $0.346 \pm 0.006$ \\ % & $0.247 \pm 0.004$ \\ 
    \rowcolor{lightblue!50!white}  ${C}_{2,0}$ & $0.6667$ & $0.5098$ & $0.4444$ & $0.3939$ & $0.3333$ \\[0.05cm]
    $\widehat{O}^{w-}_{j}$ & $0.174 \pm 0.007$ & $0.131 \pm 0.003$ & $0.114 \pm 0.002$ &  $0.100 \pm 0.002$ & $0.0844 \pm 0.0007$ \\ % & $0.063 \pm 0.020$ \\
    $\widehat{O}^{t-}_{j}$ & $0.1664 \pm 0.0003$ & $0.126 \pm 0.001$ & $0.110 \pm 0.001$ & $0.0972 \pm 0.0009$ & $0.0823 \pm 0.0005$ \\
    $(n^+_j + n^-_j)^-$ & $0.168 \pm 0.001$ & $0.1282 \pm 0.0007$ & $0.1118 \pm 0.0006$ & $0.0992 \pm 0.0006$ & $0.0840 \pm 0.0005$ \\ % & $0.0608 \pm 0.0008$ \\ 
     \rowcolor{lightblue!50!white}  ${C}_{1,0}$ & $0.1667$ & $0.1274$ & $0.1111$ & $0.09849$ & $0.08333$ \\[0.05cm]
    $s_j$ & $0.1241 \pm 0.0007$ & $0.1240 \pm 0.0008$ & $0.1240 \pm 0.0008$ & $0.1240 \pm 0.0008$ & $0.1244 \pm 0.0005$ \\ % & $0.1351 \pm 0.0004$ \\
    \rowcolor{lightblue!50!white}  $\sigma_1,\,\sigma_2$ & $0.125$ & $0.125$ & $0.125$ & $0.125$ & $0.125$ \\[0.05cm]
    $s'_j$ & $1.17 \pm 0.04$ & $1.159 \pm 0.028$ & $1.158 \pm 0.029$ & $1.16 \pm 0.03$ & $1.16 \pm 0.04$ \\ % & $1.17 \pm 0.04$ \\ 
    \rowcolor{lightblue!50!white}  $\sigma'_1,\,\sigma'_2$ & $1.125$ & $1.125$ & $1.125$ & $1.125$ & $1.125$ \\
     \bottomrule 
    \end{tabular}
       \caption{\small Scaling dimensions of lattice operators $\widehat{O}^\pm_j$  (cf. Eqs.~\eqref{O1Delta}-\eqref{eq:flip}) measured with DMRG in the $L=40$ integrable Rydberg ladder with periodic boundary conditions. The scaling dimensions of the corresponding CFT fields are highlighted.}
    \label{tab:rydoperators}
\end{table*}

\section{Critical transitions and a bubble}
\label{sec:critical}

\begin{figure*}
    \centering
    \includegraphics[width=0.96\textwidth]{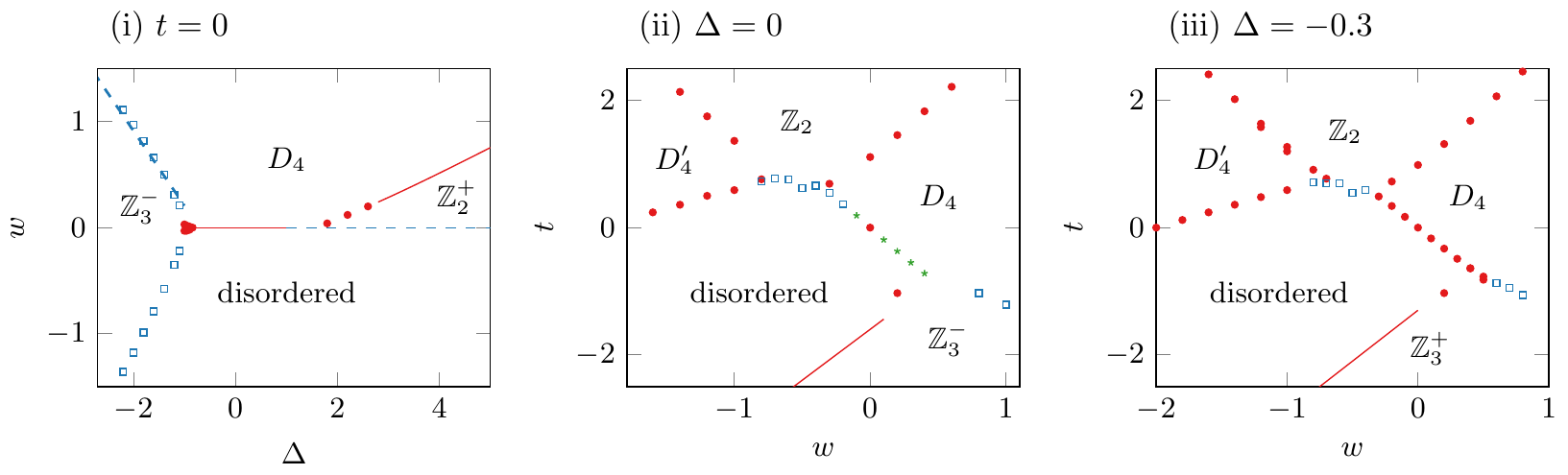}
    \caption{The phase diagram of the Rydberg ladder at (i) $t = 0$, (ii) $\Delta = 0$ and (iii) $\Delta = -0.3$ with \textcolor{darkred}{second order transitions (red filled circles)}  and \textcolor{darkblue}{first order transitions (darkblue empty squares)} and \textcolor{darkgreen}{weakly first order transitions (darkgreen stars)}. }
\label{fig:phasediagramstDelta}
\end{figure*}

In this section we do the last of the analysis needed to understand the full three-dimensional phase diagram. Our results are summarized in the three planar regions displayed in Fig.~\ref{fig:phasediagramstDelta}, along with Fig.~\ref{fig:diagramw0} above.

%
%\subsection{The effective field theory}

\subsection{The critical $D_4$-disorder transition}

The map to the XXZ chain guarantees that along the integrable $w$\,=\,$t$\,=\,0 line, the system is critical for $|\Delta|\le 1$. The nearest gapped phases to it are the $D_4$-broken and the disordered phases, and so it is natural to guess that this critical line governs a transition between the two. To explore the issue further, we analyse the effective field theory valid for $w$ and $t$ small. This field theory can be described as a perturbation of the CFT by relevant operators invariant under the lattice symmetries. The correspondences in section \ref{sec:CFT} make finding these operators simple. 
Lattice translation sends $C_{l,m}\to  (-1)^{l+m} C_{l,m}$ in the CFT, while only the operators $C_{l,0}$ are invariant under the non-invertible symmetry $\mathcal{Q}$ \cite{fendley2023pentagon}. Thus only operators $C_{2n,0}$ of dimension $n^2/R^2$ appear in the effective continuum Hamiltonian
\begin{align}
H_{\rm CFT} + \int dx \big(\lambda_1 C_{2,0}(x) + \lambda_2 C_{4,0}(x) + \kappa C_{6,0}(x)\big).
\label{Hpert}
\end{align}

On the integrable line the system possesses a self-duality under which $C_{2,0}$ and $C_{4,0}$ are not invariant. Thus the integrable line must correspond to setting $\lambda_1$\,=\,$\lambda_2$\,=0. We indeed identified $C_{2,0}$ above as the leading continuum piece of the anti-self-dual lattice perturbations $\widehat{O}^w$ and $\widehat{O}^t$. Since $C_{4,0}$ is invariant under the same lattice symmetries, it must also appear in \eqref{Hpert}. However, $C_{6,0}$ is self-dual, so $\kappa\ne 0$ even on the integrable line. This operator is of dimension $9/R^2$, so it is irrelevant for $R$\,$<$\,3/$\sqrt{2}$. It thus can be ignored until it causes a KT transition to the integrable gapped $\mathbb{Z}_2$/disorder transition line at $\Delta$\,=\,1 \cite{Braylovskaya2016}.

Taking $w$ and/or $t$ nonzero can drive the system into the $D_4$-broken or the disordered phase. For $\sqrt{2}$\,$<$\,$R$\,$<$3/$\sqrt{2}$ (--0.17\,$< \Delta\le$\,1), both $C_{2,0}$ and $C_{4,0}$ are relevant, so the transition between $D_4$ and disordered phases here should be first order and direct away from $w$\,=\,$t$\,=\,0. Our numerics indicate that the transition is weakly first-order, in that for our system sizes it still exhibits characteristics of a second-order phase transition. We denote these points accordingly in Fig.~\ref{fig:phasediagramstDelta}(ii).

For $1/\sqrt{2}$\,$<$\,$R$\,$<$\,$\sqrt{2}$ (--0.94\,$<$\,$\Delta$\,$<$\,--0.17), only $C_{2,0}$ is relevant. Perturbing by either  $\widehat{O}^w_j$ or $ \widehat{O}^t_j$ still gaps the system. However, 
since both $\widehat{O}^w_j$ and $ \widehat{O}^t_j$ renormalize onto $C_{2,0}$, there must be some linear combination of lattice couplings that makes $\lambda_1$ in \eqref{Hpert} vanish, i.e.\ $\widetilde{w}\widehat{O}^w_j$\,+\,$\widetilde{t} \widehat{O}^t_j \to C_{4,0}$ in the continuum. Perturbing by this combination thus should preserve the criticality when $C_{4,0}$ is irrelevant, and the orbifold critical line is extended to the 2d region ${w}/{t} \approx \widetilde{w}/\widetilde{t}$, for small enough $w$, $t$. 

\begin{figure}[h]
    \centering
    \includegraphics[width=0.48\textwidth]{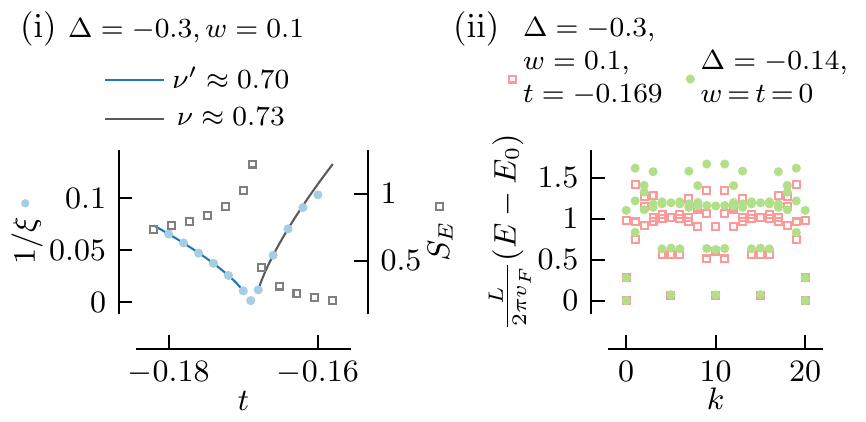}
       \caption{(i) Correlation length and half-cut entanglement entropy across the $D_4$ to disordered transition at $\Delta$\,=\,--0.3, $w$\,=\,0.1 with $L$\,=\,601, (ii) energy spectra versus momentum with $L$\,=\,20 at two transition points with a similar CFT radius $R$: 
    $(\Delta,\,w,\,t)$ = (--0.14,\,0,\,0) and (--0.3,\,0.1,--0.169).  
    }
    \label{fig:z4short}
\end{figure}

Our numerics indeed indicate a direct and critical transition occurs between $D_4$-ordered and disordered phases governed by the orbifold CFT over a two-dimensional region  with non-zero $w$ and $t$. Results for $\Delta$\,=\,--0.3 are plotted in Fig.~\ref{fig:z4short}, with the corresponding lines shown in the phase diagram in Fig.~\ref{fig:phasediagramstDelta}(iii). In our DMRG numerics, we chose system sizes $L = 12n+1$, $n \in \mathbb{N}$ for open boundary conditions so that there is a unique ground state in the $\mathbb{Z}_3$ ordered phase and two (instead of four) ground states in the $D_4$ ordered phases. The truncation error was fixed at $10^{-11}$, and the energy tolerance at $10^{-9}$ as a convergence criterion. For $L=601$, MPS bond dimensions up to 300 were typically enough to reach convergence. 

On the other hand, for 0\,$<$\,$R$\,$<$1/$\sqrt{2}\ $ (--1$<\Delta<$--0.94) all the perturbations in \eqref{Hpert} are irrelevant and the critical line is stable. A small critical cone-shaped region thus occurs in between the $D_4$, $\mathbb{Z}_3^-$ and disordered phases.  A similar critical triangle also appears in the phase diagram of the quantum Ashkin-Teller model \cite{kadanoff1981}. As we review in the Appendix, orbifolding a 4-state height model version of the XXZ chain yields the Ashkin-Teller model, which then hosts a non-invertible $U(1)$ symmetry akin to \eqref{Qdef}. %under which only the CFT operators $C_{0,n}$ are invariant. 

\begin{figure}[ht]
\centering
\includegraphics{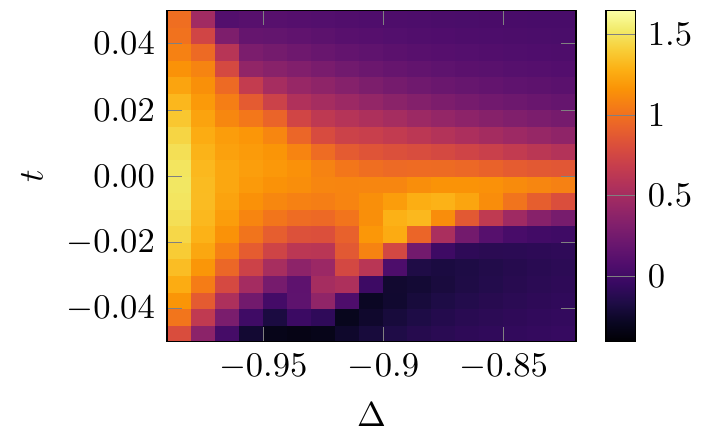}
\caption{Central charges close to $\Delta$\,=\,--1 at $w$\,=\,0 found with $L$\,=\,121 DMRG by fitting the entanglement entropy to \eqref{ccformula}.}
\label{fig:bubble}
\end{figure}

We confirm the presence of the critical bubble close to $\Delta$\,=\,--1 by numerically finding the entanglement entropy $S_E(l)$ for open boundary conditions across bond $l = 1, \dots, L/2$. In a CFT,  $S_E(l)$ obeys the Calabrese-Cardy formula  \cite{calabrese2004entanglement}
\begin{align}\label{ccformula}
S_E(l) = \frac{c}{6} \log\Big(\frac{ 2L}{\pi} \sin\frac{\pi l}{L} \Big) + {\rm const}\ . 
\end{align}
with $c$ the central charge. Convergence of DMRG is slow because $v_F\to 0$ from \eqref{Evf} as $\Delta\to$\,--1,  but we were able to find convincing evidence that the curve fits the formula throughout the triangular regions in Figs.~\ref{fig:diagramw0} and \ref{fig:phasediagramstDelta}(i). We plot the extracted central charge at $w$\,=\,0 in Fig.~\ref{fig:bubble}.  

\subsection{The chiral $\mathbb{Z}_3$ transition}
\label{sec:chiral}

The last issue we address is the nature of the $\mathbb{Z}_3^+$ to disordered transition here. This transition is apparent in Figs.~\ref{fig:diagramw0} and \ref{fig:phasediagramstDelta}(iii), with the location given approximately by \eqref{t3dis}. It is critical in chain limit of Section \ref{sec:chain}, and our numerics indicate it remains so until it collides with the first-order line separating the two $\mathbb{Z}_3$ phases. However, determining the precise type of transition requires a careful analysis, as both chiral transitions and intermediate incommensurate regions occur in the chain \cite{Fendley2003,samajdar2018,chepiga2019,giudici2019}. 

\begin{figure}
    \centering
    \includegraphics[width=0.48\textwidth]{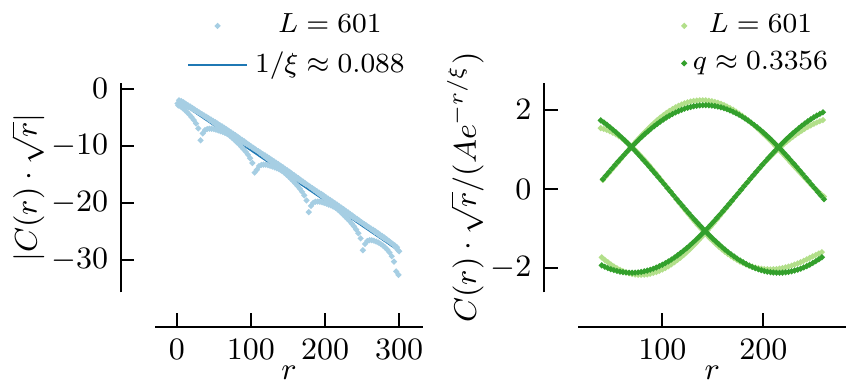}
        \caption{On the left, the logarithm of $|G(r) \sqrt{r}|$ is fitted linearly to extract the correlation length $\xi$ at $w=0$, $\Delta = -0.92$, $t = -0.63$. The reduced correlation function $G(r) \sqrt{r}/(A e^{-r/\xi})$ is then fitted to a cosine function in the right plot to extract the wave vector $q$.  }
    \label{fig:helixes3}
\end{figure}

\begin{figure}
    \centering
    \includegraphics[width=0.48\textwidth]{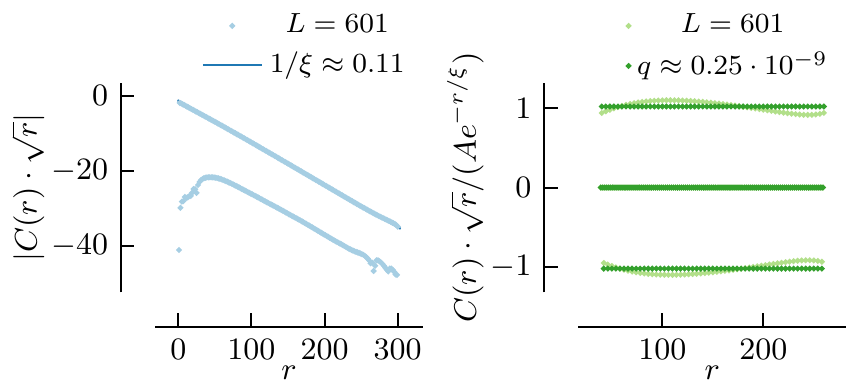}
    \caption{As with Fig.~\ref{fig:helixes3}, but here close to the $D_4$ to disorder transition at $w=0.1$, $\Delta = -0.48$, $t = -0.16$. The resulting wave vector is $q=1/4$. }
    \label{fig:helixes4}
\end{figure}

We utilize the numerical method of \cite{chepiga2019} and study the boson-density two-point correlator
\begin{align}\label{eq:correlator}
G(r) = \langle (1-n^e_r)(1-n^e_0)\rangle - \langle1-n^e_r\rangle \langle 1-n^e_0\rangle\ .
\end{align}
We use their two-step fitting procedure to fit it to the Ornstein-Zernicke relation 
\begin{align}
G(r) \propto {e^{-r/\xi}}\frac{\cos(2\pi q r + \phi_0)}{\sqrt{r}}\ . 
\end{align}
As illustrated in Figs.~\ref{fig:helixes3} and \ref{fig:helixes4}, we first extract the correlation length $\xi$ from fitting $|G(r) \sqrt{r}|$, and then use it to extract the wave vector $q$ by fitting the cosine function. 
The latter figure provides a useful check, as we do not expect the $D_4$ to disorder transition to be chiral (as observed in e.g.~\cite{maceira2022conformal} for a $\mathbb{Z}_4$ transition). The reason is that here operators $C_{n,-n}$ which induce chiral perturbations are forbidden by the non-invertible $U(1)$ symmetry $\mathcal{Q}$. We indeed find that the wave vector is always $q=1/4$, excluding the possibility of a chiral transition.

\begin{figure}
    \centering
    \includegraphics{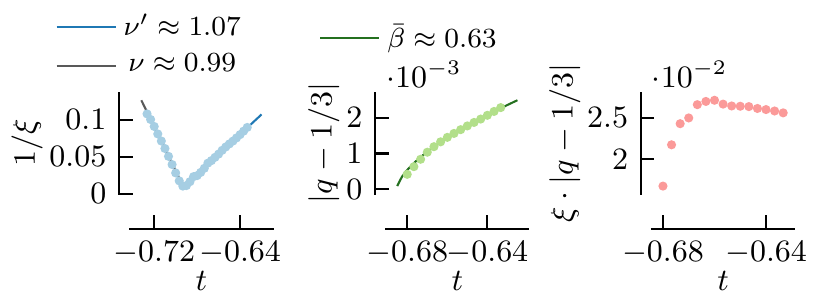}
    \includegraphics[]{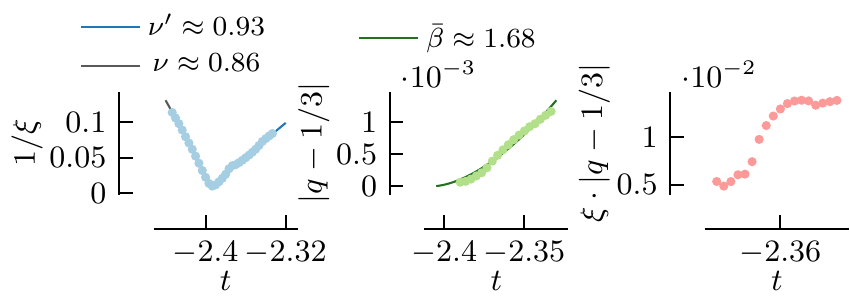}
    \caption{The wave vector $q$, correlation length $\xi$ and their product across the $\mathbb{Z}_3^+$-disorder transition at $w = 0$, $\Delta = -0.6 + 0.5t$ (top row) and $\Delta = 2 + 0.5t$ (bottom row). The DMRG data obtained with $L=601$ sites is plotted together with power-law fits for the exponents $\nu$, $\nu'$ and $\Bar{\beta}$. 
    }
    \label{fig:z3short}
\end{figure}

However, we have strong evidence that the $\mathbb{Z}_3^+$ to disordered transition indeed can be chiral. We display our results for several points along this line in Fig.~\ref{fig:z3short}. 
At small $\Delta$ (top row), the product $\xi \cdot |q - 1/3|$ seems to go to a very small but finite value and the exponent $\Bar{\beta}$ is closer to the prediction $\Bar{\beta} = \nu$ for a chiral transition than to the Potts value $\Bar{\beta} = 5/3$. 
At large $\Delta$ (bottom row), the product $\xi \cdot |q - 1/3|$ seems to go zero and the exponent $\Bar{\beta}$ is in good agreement with the Potts value $\Bar{\beta} = 5/3$. We conclude that the $\mathbb{Z}^+_3$ transition in the ladder is likely to be a chiral transition for smaller $\Delta$, and a Potts transition for large $\Delta$. We were not able to locate a sharp transition between the chiral regime and the Potts regime, and so presume that the strength of the chiral perturbations gradually decays with increasing $\Delta$.

We find no evidence of any incommensurate phase, despite one occurring in the chain. However, we note that the incommensurate phase in the chain (established definitively in an extreme limit) is rather small and impossible to see numerically. Thus our lack of evidence for its existence here is not definitive.

\section{Conclusions}

We obtained the essentially complete three-parameter phase diagram for a quantum Hamiltonian describing Rydberg bosons on a square ladder with \zz symmetry and a one-particle-per-square constraint. Various limits allowed us to find all the density-wave phases, while integrability, CFT and numerics allowed us to glue them together. Our results thus provide meaningful progress in understanding strong-coupling Rydberg-blockade physics, showing the problem is not intractable via analytic methods.

We also found a variety of interesting critical transitions, including a chiral one. Thus such transitions occur in this ladder as they do in the chain.  We find free-boson orbifold and Ising$^2$ critical lines as well, giving support to the hope \cite{slagle2022quantum} that the criticality can survive in coupled chains without excessive fine-tuning. 

Our model provides not only potentially experimentally relevant results, but also an interesting application of non-invertible symmetries.
The self-duality of the integrable line helps us find the effective theory there and in the nearby region, while the non-invertible symmetry $\mathcal{Q}$ is present throughout the phase diagram. The latter results in two striking features. One is the presence of two distinct $\mathbb{Z}_3$ density-wave orders, one where $\mathcal{Q}$ is spontaneously broken. The other is that invariance under $\mathcal{Q}$ forbids perturbing the integrable line by a number of relevant operators. This symmetry thus enhances the stability of the $D_4$-disordered transition under perturbation in one direction, resulting in a small critical cone for $\Delta$ near $-1$ and small $w,\,t$.

\section*{Acknowledgments} P.F.\ thanks Jason Alicea for many conversations on the Rydberg blockade. This work has been supported by the EPSRC Grant no. EP/S020527/1.

%\vfill\eject

\vspace{0.5cm}

\appendix

\section{Perturbation theory calculations}
\label{app:pert}

Here we give a little more detail on the perturbation theory we use to correct the location of the phase transitions found by going to various extreme limits.

\subsection{Ising\texorpdfstring{$^2$}{-squared} transition between \texorpdfstring{$D_4$}{D4} and \texorpdfstring{$\mathbb{Z}_2$}{Z2} order}

The Ising$^2$ Hamiltonian arises by neglecting creation and annihilation of $+$ bosons, the first term in \eqref{HRyd}. Because the Ising Hamiltonian can be mapped on to a free-fermion model, its phase transitions can be located exactly. In our conventions, they are where the coefficient of the $n_j^-$ term is half the magnitude of the $s_{j-1}s_{j+1}$ term. The best way to find the leading-order effect of the creation/annihilation term is thus to compute how it renormalizes the couplings while remaining in the Ising$^2$ Hilbert space. Acting with $p_j$ moves out of this space, increasing the energy by $2(\Delta+t)$. Returning to the original Hilbert space by acting with $p_j^\dagger$ then results in an extra term $-  p^\dagger p_j(1-w)^2/(2(\Delta+t))$ in the Hamiltonian. In this restricted Hilbert space, $p^\dagger p_j = (1-n^-_j)$. No other terms occur at order $(\Delta +t)$, so \eqref{eq:Hisingsquared} is corrected to
\[
 \sum_{j=1}^L \left( (1+w) s_{j-1} s_{j+1} + \left(\Delta - 2t + \frac{(1-w)^2}{2(\Delta+t)}\right) n^-_j \right)\ . 
\]
The transition thus happens at \eqref{eq:pertresult}.

To see how close the transition is this perturbative result , we use the level-crossing method (also used in e.g. \cite{Fendley2003}) to locate the Ising transition with exact diagonalization. The energy gap close to the transition is computed for $L$ and $L-4$ and it is checked at which $t$ the curves $L (\Delta E)_L$ and $(L-4) (\Delta E)_{L-4}$ cross for a given $w$ and $\Delta$. 
The numerical results for the locations of the Ising$^2$ transitions and the perturbation theory results \eqref{eq:pertresult} are plotted together in Fig.~\ref{fig:isingnew} for $w$\,=\,0. 

\subsection{First order transition between the \texorpdfstring{$\mathbb{Z}_3$}{Z3} phases}

The correction to the first-order transitions we study is found simply by computing the corrections to the ground-state energies in the corresponding extreme limit.
For $\Delta+t$ large and positive, the leading contributions to the energies of the $\mathbb{Z}_3^+$ ground state $\ket{+ee+ee\dots}$ and $\mathbb{Z}_3^-$ ground state $\ket{-ee-ee\dots}$ are
\begin{align}
\label{Z3gs}
    \epsilon_{0}^{3^+} = \frac{2L}{3} (\Delta + t)\ ,\qquad \epsilon^{3^-}_0=L\Delta\ .
\end{align}
The  $\mathbb{Z}_3^-$ ground states are exact, so they do not receive perturbative corrections. 
The leading correction to the $\ket{+ee+ee\dots}$ is at second order through the annihilation and subsequent creation of a plus boson, 
\begin{equation}
    \epsilon_2^{3^+}
= \frac{L}{3} \frac{(1-w)^2}{2(\Delta + t)},
\end{equation}
Hence the energy difference between the $\mathbb{Z}_3^+$ and $\mathbb{Z}_3^-$ ground states is 
\begin{equation}
 \epsilon_0^{3^+}  + \epsilon_2^{3^+} - \epsilon_0^{3^-} = \frac{L}{3} \left( \Delta - 2t - \frac{(1-w)^2}{2(\Delta + t)} \right). 
\end{equation}
The transition occurs when this difference vanishes up to corrections of order $(\Delta+t)^2$, yielding the first of \eqref{t3minus}.

\subsection{First-order transition between \texorpdfstring{$\mathbb{Z}_3^-$}{Z3}  and \texorpdfstring{$\mathbb{Z}_2$}{Z2} phases}

The $\mathbb{Z}_3^-$ and $Z_2$ phases both can occur as  $\Delta\to-\infty$ and $t\to+\infty$. In this limit, the $\mathbb{Z}_2$-phase ground state is $\ket{-e-e\dots}$, with zeroth order energy
\begin{align}
    \epsilon_0^{2^-} = \frac{L}{2} (\Delta - 2t).
\end{align}
The first order correction to the energy vanishes, but the second-order correction due to flipping a minus boson to a plus boson and back is given by
\begin{equation}
    \epsilon_2^{2^-} 
= \frac{L}{2} \frac{(1+2w)^2}{2(\Delta - 2t)},
\end{equation}
The energy difference between this ground state and the  ground state of the $\mathbb{Z}_3^-$ phase is therefore
\[
    \epsilon_0^{3^-} -\big(\epsilon_0^{2^-}+\epsilon_2^{2^-}\big)= 
    L \Bigg(\frac{\Delta}{2} + t - \frac{(1+2w)^2}{4(\Delta - 2t)}\Bigg).    
\]
Requiring this vanish yields the second part of \eqref{t3minus} for  $t\approx -\Delta/2$. In fact, we find that the formula holds even when $|\Delta|$ becomes order one, as apparent in Fig.~\ref{fig:diagramw0}. Pushing it even further, setting $t$\,=\,0 we obtain $w\approx (\sqrt{2}\Delta-1)/2$, which agrees well with the numerical results in Fig.~\ref{fig:phasediagramstDelta}(i).

\subsection{First-order transition between the \texorpdfstring{$\mathbb{Z}_2$}{Z2} ordered and disordered phases}

The spontaneously broken self-duality along the integrable $w$\,=\,$t$\,=0 line for $\Delta > 1$ results in a first-order phase transition between the $\mathbb{Z}_2$ ordered phase and the disordered phase. We here show how the location of this transition changes for non-vanishing $w$ and $t$. We assume both $\Delta-2t$ and $\Delta+t$ large and positive so that the off-diagonal terms in $H$ in the $\pm$ basis are small relative to the diagonal terms.  In this limit, the $\mathbb{Z}_2$ ground states are $\ket{+e+e\dots}$ and its translation, while the disordered ground state is $\ket{e\,e\,e\,\dots}$. Both have energy zero at leading order. Second-order corrections to the latter are
\begin{align}
\epsilon^{\rm dis}_2 = -L \frac{(1-w)^2}{2(\Delta + t)}\ .
\label{edis}
\end{align}
while contributions to the former come from both off-diagonal terms
\begin{align}
    \epsilon^{+}_2 = -\frac{L}{2} \frac{(1+2w)^2}{2(\Delta - 2t)} - \frac{L}{2} \frac{(1-w)^2}{2(\Delta + t)}\ .
    \label{e2plus}
    \end{align}
At the first-order transition $\epsilon^{\rm dis}_2 = \epsilon^{2^+}_2$ yielding 
\begin{align}
    \frac{(1+2w)^2}{\Delta - 2t} =\frac{(1-w)^2}{\Delta + t}\ .
\label{disz2}
\end{align}
Solving for $t$ gives \eqref{eq:firstorderresult}.

The assumption in deriving \eqref{disz2} was that all terms in $\epsilon^{\rm dis}_2$ and $\epsilon^{2^+}_2$ remain small in magnitude. However, when comparing with our numerics in Fig.~\ref{fig:firstorder}, we saw that agreement is good even when the denominators in \eqref{edis} and \eqref{e2plus} vanish. The reason is that in setting  $\epsilon^{\rm dis}_2 = \epsilon^{2^+}_2$ to obtain \eqref{disz2}, we force the numerators to vanish at these points as well. Thus at these special points ($t$, $w$)\,=\,($-\Delta$,\,1) and ($\Delta/2$,\,$-1/2$), the values of each side of \eqref{disz2} are still order 1/$\Delta$  (they are $3/\Delta$ and $3/(2\Delta)$ respectively). Thus at large enough $\Delta$, the curve \eqref{eq:firstorderresult} should still accurately describe the location of the transition. We found even $\Delta=2.5$ works reasonably well, cf. Fig.~\ref{fig:firstorder}.

\section{The quantum Ashkin-Teller model}

\subsection{Mapping to the XXZ chain}

The integrable Rydberg ladder as well as the XXZ chain can be written as 
\begin{equation}\label{eq:SP}
    H = \sum_{j=1}^L \left( S_j + \Delta P_j \right),
\end{equation}
with $S_j$ and $P_j$ satisfying the $Q = 3$ chromatic algebra enhanced by the Jones-Wenzl projector \cite{fendley2023pentagon},
%\begin{equation}
\begin{gather}\label{eq:algebra}
    P^j = P_j, \quad S^2_j = \mathbb{I} - P_j, \quad S_j P_j = P_j S_j = 0, \cr
    S_j S_{j \pm 1} S_j = P_j S_{j \pm 1} P_j = 0.
\end{gather}
%\end{equation}

The Ashkin-Teller model describes two Ising models on the same lattice coupled across each of the bonds. The quantum 
Hamiltonian is built from two sets of Pauli matrices acting on each site of the chain. We include a symmetry interchanging the two chains, so the Hamiltonian is
\begin{equation}\label{eq:H_AT}
\begin{aligned}
    H_\text{AT} = - \sum_{j=1}^L \Big( &\sigma^z_j \sigma^z_{j+1} + \beta \sigma^x_j + \tau^z_j \tau^z_{j+1} + \beta \tau^x_j \\
    &- \Delta (\sigma^z_j \sigma^z_{j+1} \tau^z_j \tau^z_{j+1} + \beta \sigma^x_j \sigma^x_{j+1}) \Big).
\end{aligned}
\end{equation}
The ensuing $D_4$ symmetry generators include both spin-flip symmetries 
\begin{align}
F_\sigma =  \prod_{j=1}^L\sigma^x_{j}\ ,\quad
F_\tau= \prod_{j=1}^L \tau^x_{j} \ ,
\end{align}
along with chain exchange.
At the self-dual point $\beta$\,=\,1, the Hamiltonian \eqref{eq:H_AT} can be split up in a similar fashion as \eqref{eq:SP}:
\begin{align}
    H_\text{AT} = - \sum_{k=1}^{2L} \left( S_k + \Delta P_k \right)\ ,
\end{align}
where the $2L$ operators 
\begin{align*}
    S_{2j} &= \tfrac{1}{2} \left(\sigma^x_j + \tau^x_j \right), \quad\  S_{2j+1} = \tfrac{1}{2} \left(\sigma^z_j \sigma^z_{j+1} + \tau^z_j \tau^z_{j+1} \right),\\
    P_{2j} &= \tfrac{1}{2} \left( \mathbb{I} - \sigma^x_j \tau^x_j \right), \quad P_{2j+1} = \tfrac{1}{2} \left( \mathbb{I} - \sigma^z_j \sigma^z_{j+1} \tau^z_j \tau^z_{j+1} \right).
\end{align*}
satisfy the same algebra \eqref{eq:algebra}. 

Since their generators obey the same algebra, we expect that there exists a non-invertible duality between the Ashkin-Teller and XXZ chains. This duality is found in the classical models by orbifolding a height-model version of the 6-vertex model with $2L$ sites \cite{fendley1989}. The height model has four heights $h_j \in \{1,2,3,4\}$ on each site $j = 1, \dots, 2L$ that obey the constraint $h_{j+1} = h_j \pm 1$. The two domain walls between adjacent heights correspond to the XXZ degrees of freedom.  
Orbifolding by the $\mathbb{Z}_2$ symmetry $2 \leftrightarrow 4$ in the height model yields the Ashkin-Teller model:
\begin{equation}
    \begin{tikzpicture}[baseline={([yshift=-.5ex]current bounding box.center)}]
        \draw[] (0,0) rectangle (1,1);
        \node[below] at (0,0) {1};
        \node[below] at (1,0) {2};
        \node[above] at (1,1) {3};
        \node[above] at (0,1) {4};
        \draw[dashed, darkblue] (0,0) -- (1,1);
        \draw[latex-latex, darkblue] (1.5,0.5) -- (3,0.5);
        \node[above] at (2.25,0.5) {orbifold};
    \begin{scope}[shift={(3.5,0)}]
        \draw[] (0,0) -- (1,1);
        \draw[] (0,1) -- (1,0);
        \node[below] at (0,0) {1};
        \node[below] at (1,0) {1'};
        \node[above] at (1,1) {3};
        \node[above] at (0,1) {3'};
        \node[right=2mm] at (0.5, 0.5) {2};
    \end{scope}
    \end{tikzpicture}
\end{equation}
The Ashkin-Teller adjacency diagram requires that every state $s_j \in \{1, 1', 3, 3'\}$ is adjacent to a state $s_{j+1} = 2$. Hence effectively, the sublattice with the $s_{2j} = 2$ states can be neglected, and the Ashkin-Teller model can be defined on the sublattice $s_{2j+1} \in \{1,1', 3, 3'\}$.
The states $1$, $1'$, $3$, $3'$ are identified with the $\sigma^z$, $\tau^z$ eigenstates in the quantum model:
\begin{equation}
\begin{aligned}
    1 &\to \ket{0}_\sigma \ket{0}_\tau, \quad 1' \to \ket{1}_\sigma \ket{1}_\tau, \\
    3 &\to \ket{0}_\sigma \ket{1}_\tau, \quad 3' \to \ket{1}_\sigma \ket{0}_\tau.
\end{aligned}
\end{equation}

This orbifold mapping from XXZ to Ashkin-Teller is equivalent to applying Kramers-Wannier duality to one sublattice \cite{kadanoff1981}. The map then can be written as a matrix-product operator, with weights given by
\begin{equation}
\begin{aligned}
     \begin{tikzpicture}[baseline={([yshift=-.5ex]current bounding box.center)}, scale = 0.7]
    \draw[dashed] (0,0) rectangle (2,1);
    \node[below] at (0,0) {$h \in \{2, 4\}$};
    \node[below] at (2,0) {$1$};
    \node[above] at (0,1) {$2$};
    \node[above] at (2,1) {$s \in \{1,1'\}$};
  \end{tikzpicture}
   &= 2^{-1/4} (-1)^{\delta_{h4} \delta_{s1}}, \\
  \begin{tikzpicture}[baseline={([yshift=-.5ex]current bounding box.center)}, scale = 0.7]
    \draw[dashed] (0,0) rectangle (2,1);
    \node[below] at (0,0) {$h \in \{2, 4\}$};
    \node[below] at (2,0) {$3$};
    \node[above] at (0,1) {$2$};
    \node[above] at (2,1) {$s \in \{3,3'\}$};
  \end{tikzpicture}
   &= 2^{-1/4} (-1)^{\delta_{h4} \delta_{s3}}.
\end{aligned}
\end{equation}
%\paul{You changed roles of $h$ and $s$. Also need to define $A,B,C,D$ and mention constraints} 
It is useful to consider the combinations $\ket{\pm}_1$ and $\ket{\pm}_3$ which are simultaneous eigenstates of $\sigma^x_j \tau^x_j$ and $\sigma^z_j \tau^z_j$: 
\begin{equation}
\begin{aligned}
    \ket{\pm}_1 &= \frac{1}{\sqrt{2}} \big( \ket{0}_\sigma \ket{0}_\tau \pm \ket{1}_\sigma \ket{1}_\tau \big), \\
    \ket{\pm}_3 &= \frac{1}{\sqrt{2}} \big( \ket{0}_\sigma \ket{1}_\tau \pm \ket{1}_\sigma \ket{0}_\tau \big).
\end{aligned}
\end{equation}
Knowing three successive heights $(h_{j-1}, h_j , h_{j+1})$ in the 4-state height model gives the state $s_j$ in the quantum Ashkin-Teller model via
%\begin{equation}\label{eq:Odefect}
\begin{align*}
    \{ \ket{BAB}, \ket{DAD} \} &\to \ket{+}_1, \; \{ \ket{BCB}, \ket{DCD} \} \to \ket{+}_3, \\
    \{ \ket{BAD}, \ket{DAB} \} &\to \ket{-}_1, \;  \{ \ket{BCD}, \ket{DCB} \} \to \ket{-}_3.
\end{align*}
%\end{equation}

When $\beta \ne 1$, applying the Kramers-Wannier transformation to the $\tau$ spins in the Ashkin-Teller model, followed by applying it to both $\sigma$ and $\tau$ spins yields the staggered XXZ chain \cite{kadanoff1981},
\begin{equation}
    H = \sum_{j=1}^L \left( 1 + (-1)^{j} \right) \left( \sigma^x_j \sigma^x_{j+1} + \sigma^y_j \sigma^y_{j+1} + \Delta \sigma^z_j \sigma^z_{j+1} \right).
\end{equation}

\subsection{Non-invertible symmetries}

We expect that as in the Rydberg ladder, the square of the $U(1)$ charge of the XXZ height model can be mapped to the Ashkin-Teller model. We find this non-invertible charge to be
\begin{equation}\label{eq:Q_AT}
\begin{aligned}
    &\mathcal{Q}_\text{AT} = \frac{1}{2} \left( 1 + F_\sigma F_\tau \right) \sum_{j=1}^L \sum_{k=0}^{L-1} (-1)^{k+1} \prod_{l=0}^k \left(  \sigma^x_{j+l} \tau^x_{j+l} \right)\ .
\end{aligned}
\end{equation}
The charge $\mathcal{Q}_\text{AT}$ is non-vanishing only in half the Hilbert space, as $F_\sigma F_\tau \mathcal{Q}_\text{AT} =  \mathcal{Q}_\text{AT}$. 

Applying Kramers-Wannier duality to both $\sigma$ and $\tau$ spins  gives rise to a duality under which $\beta \to 1/\beta$. This is reminiscent of the $w \to -w$, $t \to -t$ duality in the Rydberg ladder. 

\subsection{Ground states and symmetry breaking}

In the limit $\Delta \to -\infty$, the two ground states of \eqref{eq:H_AT} have eigenvalue $-1$ under $\sigma^x_j \tau^x_j$ and $\sigma^z_{j} \sigma^z_{j+1} \tau^z_j \tau^z_{j+1}$ for all $j$, and so are
\begin{equation}
    \ket{-_1 -_3 -_1 -_3 \dots }, \quad \ket{-_3 -_1 -_3 -_1 \dots}.
\end{equation}
These states are exact eigenstates of the Hamiltonian, and so they persist for any finite $\Delta < -1$.
They spontaneously break $\mathbb{Z}_2$ translation symmetry, and they have the maximal charge $\mathcal{Q}_\text{AT} = L$ under the non-invertible $U(1)$ symmetry \eqref{eq:Q_AT}.

For $\Delta \to \infty$, the two ground states of \eqref{eq:H_AT} instead have eigenvalue 1 under these operators, and so are
\begin{equation}
    \ket{+_1 +_1 +_1 +_1 \dots }, \quad \ket{+_3 +_3 +_3 +_3 \dots}.
\end{equation}
These two ground states are mixed by the symmetry generators $F_\sigma$ and $F_\tau$, and so the $D_4$ symmetry is thus spontaneously broken in this phase. Translation symmetry, however, is preserved. 

\subsection{Field theory of the quantum Ashkin-Teller chain}

To understand the transition between the $D_4$-broken phase and the $\mathbb{Z}_2$-broken phases here, we study the effective field theory outside of the critical regime.
Using the mapping to the staggered XXZ chain, this field theory  can be derived with the Coulomb-gas approach \cite{kadanoff1981}. We here restate these results in the CFT language.
The critical regime of the Ashkin-Teller model is described by a free-boson orbifold CFT with radius \cite{Yang1987}
\begin{equation}\label{eq:ATradius}
    \Delta = - \cos{\left(\frac{\pi r^2 }{2}\right)},
\end{equation}
which is twice the radius of the corresponding XXZ chain. In the region near its critical line, the effective field theory describing the continuum limit of the Ashkin-Teller chain in the region near the critical line can be written as a perturbed CFT. The only operators invariant under both translation and the non-invertible $U(1)$ symmetry are $C_{0,2n}$, so the resulting effective Hamiltonian is
\begin{equation}\label{eq:ATfieldtheory}
    H_\text{CFT} + \int dx \; \big( \lambda_1 C_{0,2}(x) + \lambda_2 C_{0,4}(x) \big).
\end{equation}

The critical line extends into a critical fan when both perturbations are irrelevant, as with our Rydberg-blockade ladder. Here that occurs when  $r^2$\,$<$\,1/2 ($\Delta < -1/\sqrt{2})$. At larger values of $\Delta$,  $1/2 < r^2 < 2$ ($-1/\sqrt{2} < \Delta < 1$), the operator $C_{2,0}$ is relevant and the model is critical only at $\beta$\,= 1. At $r^2$\,=\,2 ($\Delta$\,=1), the operator $C_{0,4}$ with scaling dimension $4/r^2$ becomes relevant and drives a KT transition. 

\bibliographystyle{apsrev}
\bibliography{references}

\end{document}